# Effect of He Self-organized pattern plasma-activated media with different conductivity on cancer cells

Zhitong Chen[1,2]*

[1]Department of Mechanical and Aerospace Engineering, The George Washington University, Washington, DC 20052, USA

[2]National Innovation Center for Advanced Medical Devices, Shenzhen, Guangdong 518000, China

**Abstract**

The self-organized pattern (SOP) phenomenon is prevalent in plasma, while knowledge about SOP discharge affecting reactive species generated plasma-activated media (PAM) for cancer therapy is poorly documented. The aim of this study focused on the effect of SOP discharge modes on reactive oxygen and nitrogen species (ROS, RNS) in He SOP plasma-activated media with different conductivity (saline solution and deionized (DI) water), and employed them to breast cancer MDA-MB-231 and pancreatic BxPC-3 cancer cells. Optical emission spectrum and Fluorimetric analysis were used to identify and quantify ROS and RNS generated in He SOP plasma-activated saline solution and DI water. Furthermore, He SOP plasma discharge modes are capable of efficiently controlling the ROS and RNS concentration in the plasma-activated saline solution and DI water, which contribute to the cytotoxic effect. On the other hand, stainless steel and copper were used as a lower electrode to compare their effect on cell viability. Taken together, our findings provide insight into potential mechanisms involved in cell death after treatment with He SOP plasma-activated media.

* Corresponding Author:
E–mail address: zt.chen@nmed.org.cn

## 1. Introduction

Patterns are coherent structures or arrangements with a clear degree of temporal, spatial or spatiotemporal regularity, and their processes are constituted by sequences of events leading to the eventual establishment of a pattern[1]. Such event often occurs in a self-organized manner in nature without structured interventions or active manipulation of external constraints. As in many physical, chemical, and biological systems, the self-organized pattern phenomenon is also prevalent in plasma[2]. Various types of pattern formation phenomena have been reported in a wide range of plasmas, such as dielectric barrier discharges (DBDs), arc discharges, high-pressure-low-current glow discharges, high-pressure-high-current-arc discharge, low-pressure-low-current glow, and low-pressure-high-current vacuum arc discharge[3-7]. The self-organized pattern is also often found in plasmas interacting with liquid surfaces, of relevance in applications ranging from water decontamination and activation, to material synthesis, and medicine[8-11]. The self-organizational restructuration and patterning at the interfaces drastically widen the spectrum of the involved structural, physical and chemical processes, thus opening new horizons to tackling plasma media for medical applications. On the other hand, cold plasma has been introduced as a novel therapeutic method for anticancer treatment[12-15]. Moreover, plasma has shown significant potential for various biomedical applications such as inactivation of microorganisms, wounding healing, sterilization of infected tissues, blood coagulation, skin regeneration, tooth bleaching and cancer therapy[16-24]. The efficacy of plasma in the proposed applications relies on the synergistic action of the reactive oxygen species (ROS) and reactive nitrogen species (RNS)[3,25]. Thus, we put forward the idea that helium self-organized pattern (He SOP) plasma-stimulated saline solution and deionized water enriched with ROS, RNS, and other active species applied the tumor cells. Discharge modes amendable for adaptation of plasma-stimulated media were proposed but yet realized. In this work, we present a novel approach to study He SOP plasma discharge modes and their effect on the cancer-killing plasma-activated therapeutic media. After a short treatment of He SOP with different discharge modes, the cancer-inhibiting media has acquired a pronounced cancer-suppressing activity towards at least two kinds of human cancer cells, namely breast cancer MDA-MB-231 and pancreatic BxPC-3 cancer cells.

## 2. Experimental design

*2.1 He SOP plasma device*

Figure 1 shows a schematic representation of the AC power and He SOP plasma discharge setup. The discharge modes and self-organized patterns were organized as follows. The lower electrode (thin copper/stainless steel plate, thickness d = 0.2 mm, Ø = 22 mm) was placed at the bottom of a treat well. Above the plate, 6 ml of saline solution and deionized water were added to the well. The upper electrode (inner diameter Ø = 4 mm) was then installed above the liquid surface. AC power supply unit was fabricated. Voltage is applied between two electrodes, and the gap between the electrode and liquid surface accommodated a bunch of plasma. According to discharge modes of saline solution and DI water, we have selected flow rate at 0.6 L/min as basic platforms for the bio-oriented studies. SOP plasma discharge setup is capable of producing well-defined self-organized patterns between one electrode and the liquid media surface. Saline solution and DI water were treated by He SOP plasma with 90 seconds' duration at 6V, 8V, 10V, and 12V to obtain plasma solutions applied to cancer cells.

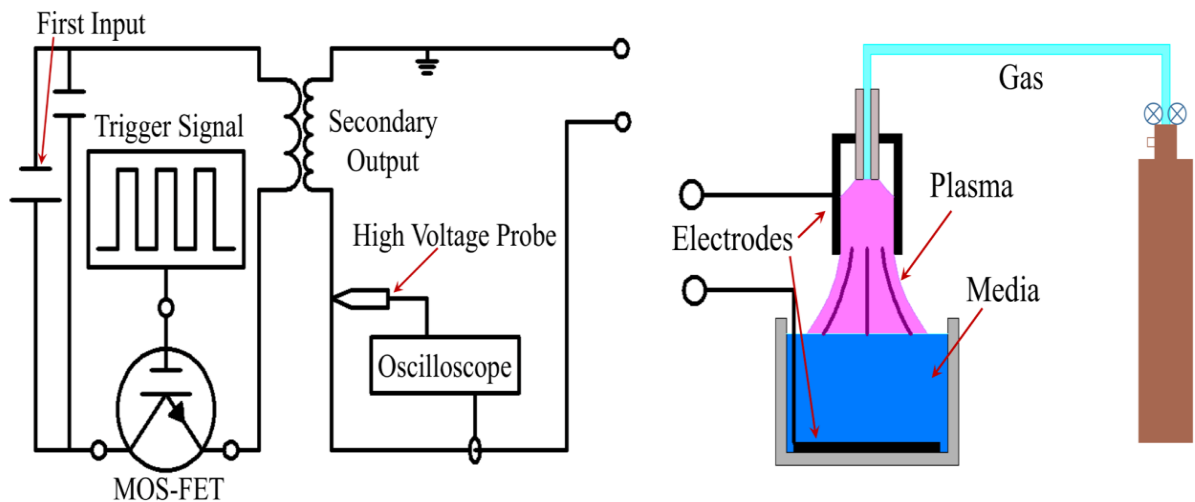

Figure 1 Schematic representation of AC power and the He SOP plasma discharge setup. Around 10 mm gap between the electrode and surface of liquid accommodated a bunch of plasma.

*2.2 Optical emission spectra measurement*

UV-visible-NIR, a range of wavelength 200-850 nm, was investigated on plasma to detect various RNS and ROS (nitrogen [$N_2$], nitric oxide [–NO], nitrogen cation [$N^{+2}$], atomic oxygen [O], and hydroxyl radical [–OH]). The spectrometer and the detection probe were purchased from Stellar Net Inc. In order to measure the radius of the He SOP plasma-activated saline solution and DI water, the optical probe was placed at a distance of 2 cm in front of the plasma. The integration time of the collecting data was set to 100 ms.

*2.3 Cell culture*

Cells (MDA-MB-231) were cultured in Dulbecco's Modified Eagle Medium (DMEM, Life Technologies) supplemented with 10% (v/v) fetal bovine serum (Atlantic Biologicals) and 1% (v/v) penicillin and streptomycin (Life Technologies). The human pancreas adenocarcinoma cancer cell line (BxPC-3) was acquired from American Type Culture Collection (ATCC). Cell lines were cultured in RPMI-1640 Medium (ATCC® 30-2001™) supplemented with 10% (v/v) fetal bovine serum (Atlantic Biologicals) and 1% (v/v) penicillin and streptomycin (Life Technologies). Cultures were maintained at 37 °C in a humidified incubator containing 5% (v/v) $CO_2$.

*2.4 Evaluation of $H_2O_2$ concentration*

Fluorimetric Hydrogen Peroxide Assay Kit (Sigma-Aldrich) was used for measuring the amount of $H_2O_2$. A detailed protocol can be found on the Sigma-Aldrich website. Briefly, we added 50 $\mu$l of standard curves samples, controls, and experimental samples (saline solution and DI water treated by He SOP plasma with input voltage 6, 8, 10, and 12V for 90 seconds) to the 96-well flat-bottom black plates, and then added 50 $\mu$l of Master Mix to each of wells. We incubated the plates for 20 min at room temperature protected from light on and measured fluorescence by Synergy H1 Hybrid Multi-Mode Microplate Reader at Ex/Em: 540/590 nm.

*2.4 Evaluation of $NO_2^-$ concentration*

RNS levels were determined by using the Griess Reagent System (Promega Corporation) according to the instructions provided by the manufacturer. Briefly, we added 50 $\mu$l of standard curves samples, controls, and experimental samples to the 96-well flat-bottom plates. Then dispense 50 $\mu$l of the Sulfanilamide Solution to all samples and incubate 5-10 minutes at room temperature. Finally, dispense 50 $\mu$l of the NED solution to all wells and incubate at room temperature 5-10 minutes. The absorbance was measured at 540 nm by Synergy H1 Hybrid Multi-Mode Microplate Reader.

*2.5 Cell viability of MDA-MB-231 and BxPC-3*

The MDA-MB-231 and BxPC-3 cancer cells were plated in 96-well flat-bottom microplates at a density of 3000 cells per well in 70 $\mu$L of complete culture medium. Cells were incubated for 24 hours to ensure proper cell adherence and stability. On day 2, 30 $\mu$L of and experimental samples (saline solution and DI water treated by He SOP plasma with input voltage 6, 8, 10, and 12V for 90 seconds), DMEM/RPMI, and saline solution/DI water were added to the cells. Cells were further incubated at 37 °C for 24 and 48 hours. The cell viability of the breast and pancreas adenocarcinoma cancer cells was measured for each incubation time point with an MTT assay. 100 µL of MTT solution (3-(4, 5-dimethylthiazol-2-yl)-2,5-diphenyltetrazolium bromide) (Sigma-Aldrich) was added to each well followed by 3-hour incubation. The MTT solution was discarded and 100 µL per well of MTT solvent (0.4% (v/v) HCl in anhydrous isopropanol) was added to the wells. The absorbance of the purple solution was recorded at 570 nm with the Synergy H1 Hybrid Multi-Mode Microplate Reader.

*2.6 Statistical analysis*

All results were presented as mean ± standard deviation plotted using Origin 8. Student's t-test was applied to check the statistical significance (*$p < 0.05$, **$p < 0.01$, ***$p < 0.001$).

## 3. Results

*3.1 He SOP plasma discharge modes*

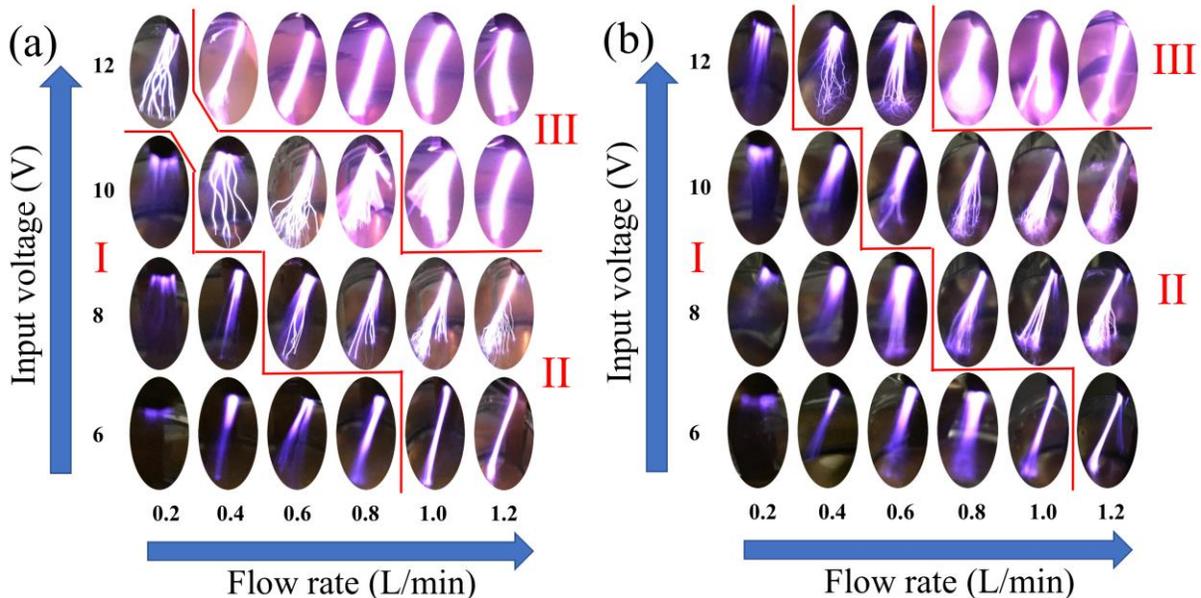

Figure 2. Optical photographs of the discharge patterns above saline solutions (a) and DI water (b) during the activation process. The self-organized patterns have complex structure strongly depending on the voltage and flow rate conditions.

Fig. 2 shows optical photographs of various self-organized discharge patterns above saline solutions and DI water. Complex shapes of discharge modes depend on the input voltage/flow rate characteristics. The input voltage/flow rate characteristics of He SOP plasma-activated saline solution and DI water can be divided into three stages. At stage I, the discharge voltage or flow rate is relatively low, and the discharge modes represent glow discharge. As the discharge voltage or flow rate increases to a certain degree, the discharge modes change from glow discharge (Stage I) to spark discharge (Stage II). Compared saline solution to DI water, He SOP activated saline solutions are easier to reach Stage II. Complex discharge modes consisting of multi-filaments and confocal lines of different densities are produced. The discharge stabilizes at the multi-filament modes and stretches out many of discharge laments between upper electrode and liquid media surface. At same conditions, He SOP plasma-activated saline solutions discharge much stronger than DI water because the saline solution has much better conductivity than DI water. Then, the discharge enters an arc discharge mode (Stage III), resulting in drastically enhanced heat radiation. High temperature results in stronger thermionic rather than secondary electron emission form a comparatively cold electrode. The discharge flips from the multi-lament and heat radiation-supported stage where the thermionic emission is enough to maintain the discharge current. Based on the above consideration, we have selected flow rate at 0.6 L/min and input voltage from 6V to 12 V for both saline solution and DI water as the basic platforms for the bio-oriented studies described below in detail.

*3.2 He SOP plasma-activated saline solution*

Fig. 3 shows optical photographs, discharge voltage, and optical emission spectra of He SOP plasma-activated saline solutions at a flow rate of 0.6L/min with different 6V, 8V, 10V, and 12V input voltages. With increasing input voltage from 6V to 8V, the discharge modes change from glow discharge to spark discharge. From 8V to 10V, the density of discharge increases, but they still are spark discharge modes. When the input voltage reaches 12V, the discharge mode becomes arc discharge, resulting in drastically enhanced heat radiation. The discharge voltage of He SOP plasma-activated saline solution at a flow rate of 0.6 L/min increased with input voltage up to 10V. While between 10V and 12V, the discharge voltage decreased, but the discharge current drastically increased. On the other hand, we have measured spectra of plasma from the plasma-liquid interface. Typical optical emission spectra are shown in Fig. 3. The emission intensity increases with the increasing input voltage, and the intensity at 12V input voltage is over the range of UV-visible-NIR. The identification of the emission bands was performed according to Pearse et al[26]. In the 250-300 nm wavelength range, relative weak emission bands (258, 267, and 297) were detected as NO lines[27]. Species at wavelengths of 337 and 357 nm were defined as $N_2$ $C^3\Pi u$ or NO $\beta^3\Pi g$ (denoted as $N_2$/NO), because both species have possible optical emission at these wavelengths[26]. The emission bands between 300 and 500 nm have still not been clearly identified in the literature[28]. However, we anticipated that OH was present at 309 nm, the wavelength of 391 nm could be indicative of $N_2^+/N_2$, and atomic oxygen (O) was denoted at the wavelength of 777 nm. The features were assigned as helium (He) lines between 550–750 nm. The dominant species of the spectra in He SOP activated saline solution were NO or N2 lines (258, 267, 297, 337, 357, and 381nm), OH (309 nm), $N_2^+$ (391 nm), and O (777 nm).

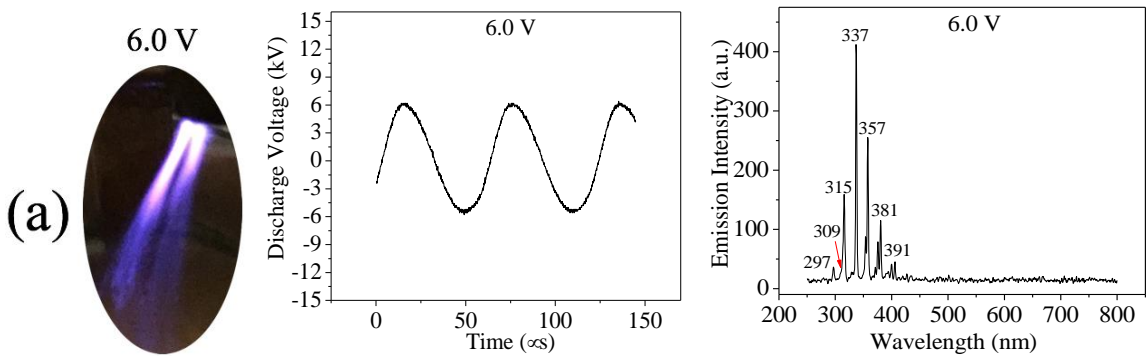

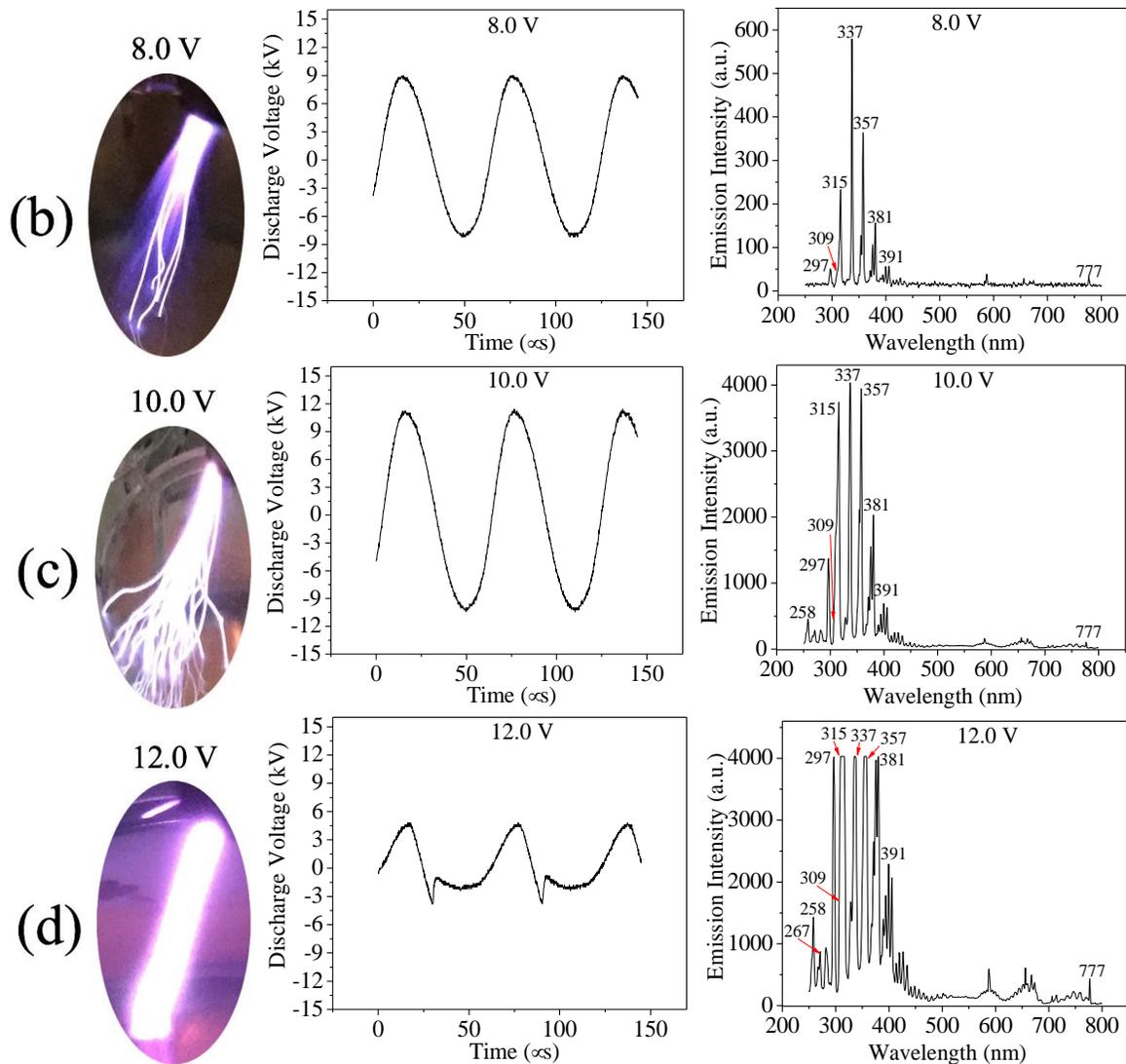

Figure 3 Optical photographs, discharge voltage, and optical emission spectra of He SOP plasma activated saline solutions at flow rate of 0.6 L/min with different input voltages: (a) 6 V, (b) 8 V, (c) 10 V, and (d) 12 V.

*3.3 He SOP plasma-activated DI water*

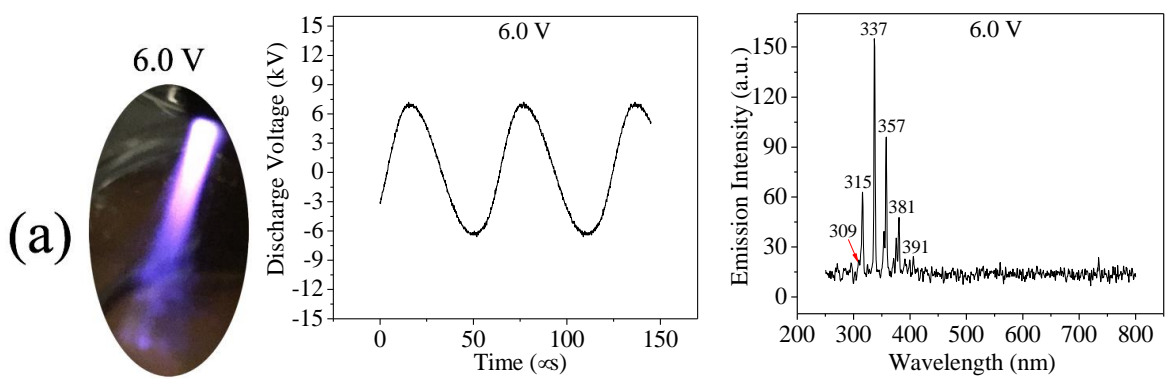

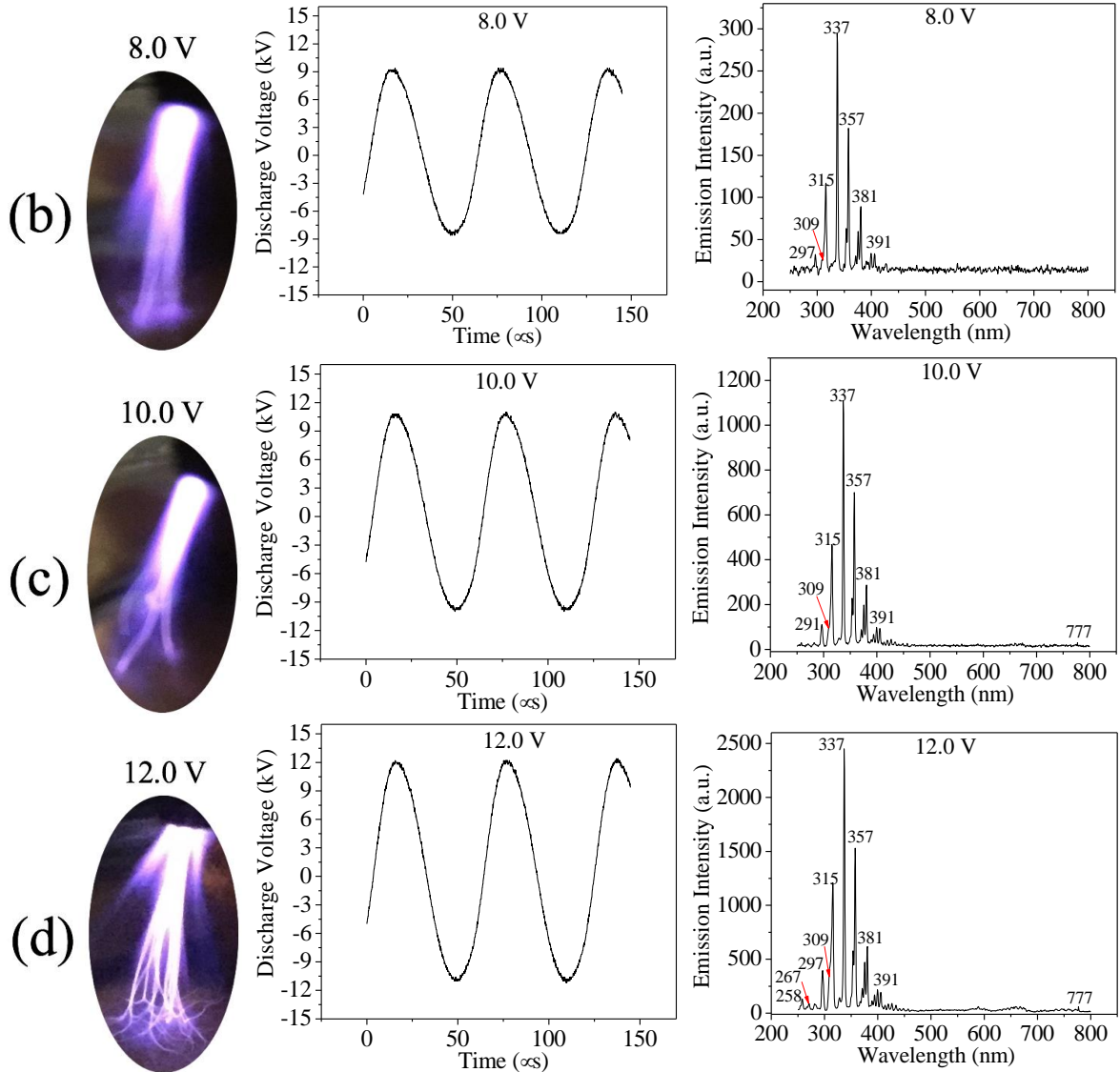

Figure 4 Optical photographs, discharge voltage, and optical emission spectra of He SOP plasma activated DI water at flow rate of 0.6 L/min with different input voltages: (a) 6 V, (b) 8 V, (c) 10 V, and (d) 12 V.

Fig. 4 shows optical photographs, discharge voltage, and optical emission spectra of He SOP plasma-activated DI water at a flow rate of 0.6L/min with different 6V, 8V, 10V, and 12V input voltages. For 6V and 8V input voltages, the discharge modes appear glow discharge; for 10V and 12V input voltage, the discharge modes appear spark discharge. The discharge voltage of He SOP plasma-activated DI water increased with input voltage. Compared He SOP plasma-activated saline solution with DI water at input voltage 12V, although discharge voltage of DI water is higher than saline solution, its current is much lower than saline solution. The optical emission spectra of He SOP plasma-activated DI water at 0.6 L/min were also measured via UV-visible-NIR. The emission intensity increased with increasing input voltage. Compared with saline solution, He SOP plasma-activated DI water has lower emission intensity. The features were assigned as helium (He) lines between 550-750 nm. The dominant species of the spectra in He SOP activated DI water were NO or N2 lines (258, 267, 297, 337, 357, and 381 nm), OH (309 nm), $N_2^+$ (391 nm), and O (777 nm).

*3.4 $H_2O_2$ and $NO_2^-$ concentration of He SOP plasma-activated saline solution and DI water*

He SOP plasma can produce chemically activated species in saline solution and DI water. He SOP plasma irradiation produced $H_2O_2$ and $NO_2^-$ in saline solution and DI water in an input voltage-dependent manner, as shown in Fig. 5. A description of the possible mechanism of $H_2O_2$ and $NO_2^-$ formation can be found in our previous publications[8,13]. Compared to saline solutions with DI water, He SOP plasma-activated saline solutions have lower $H_2O_2$ concentrations than DI water, while its $NO_2^-$ concentrations are much higher than DI water. From Fig. 3 and Fig. 4, we know that the spectra intensity of He SOP plasma-activated saline solution is much higher than DI water at each input voltage, which means $H_2O_2$ and $NO_2^-$ concentration of saline solution should be much higher than DI water. The presence of chloride ions ($Cl^-$) in saline solution may be a key reason for lower $H_2O_2$ concentration. In the plasma process, the O could lead by reaction with the presence $Cl^-$ in the liquid phase to form $OCl^-$[29]. The $OCl^-$ will react with $H_2O_2$ forming $Cl^-$, $O_2$, and $H_2O$[30]. The presence of Cl- should be critical to obtain the depletion in $H_2O_2$ concentration.

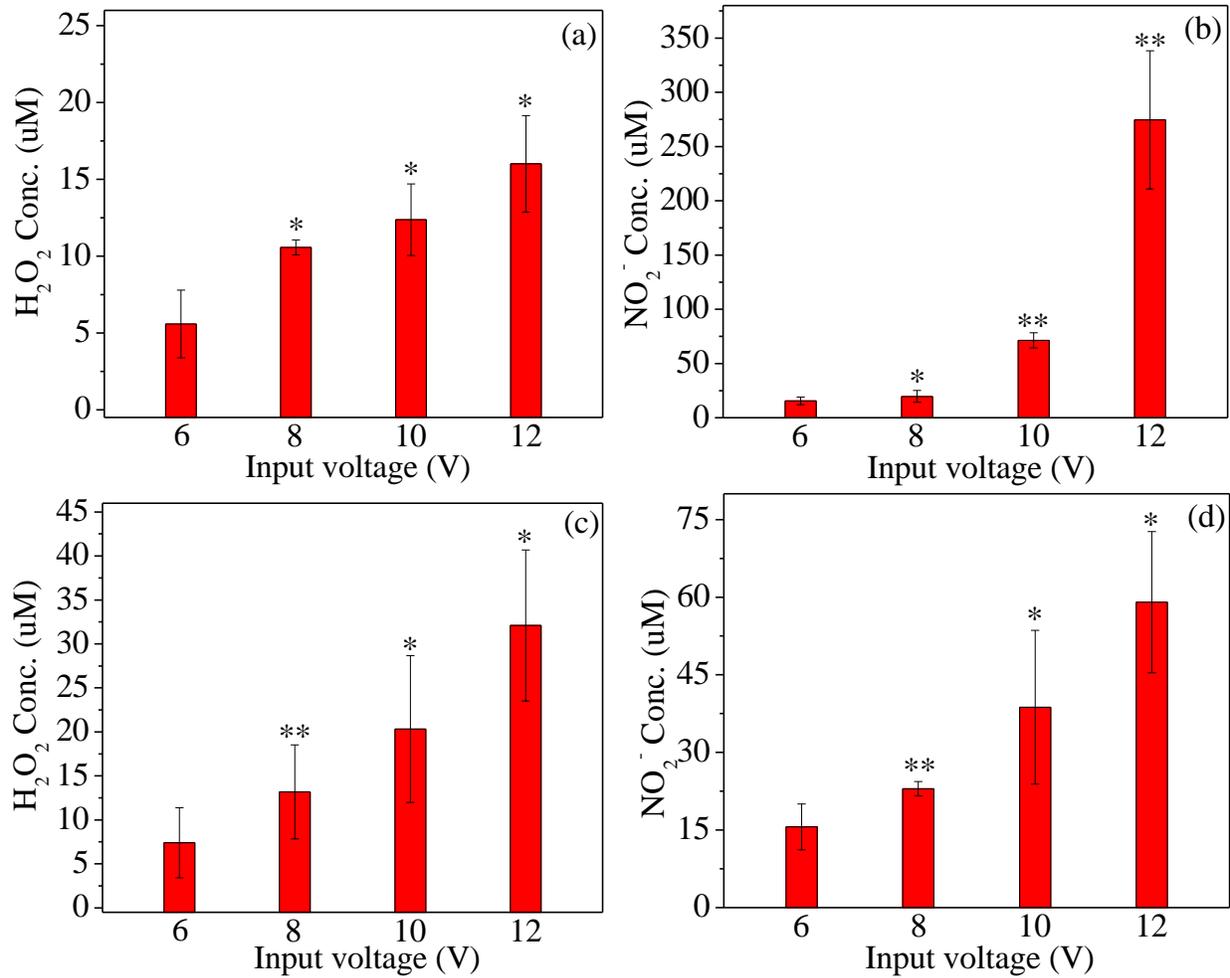

Figure 5 $H_2O_2$ and $NO_2^-$ concentration in saline solution and DI water produced by He SOP plasma in an input voltage-dependent manner. (a) $H_2O_2$ and (b) $NO_2^-$ concentration for He SOP plasma activated saline solutions; (c) $H_2O_2$ and (d) $NO_2^-$ concentration for He SOP plasma activated DI water. Student t-test was performed, and the statistical significance compared to input voltage 6 V is indicated as *$p < 0.05$, **$p < 0.01$, ***$p < 0.001$. (n=3)

*3.5 Cell viability of He SOP plasma-activated saline solution and DI suing copper as lower electrode*

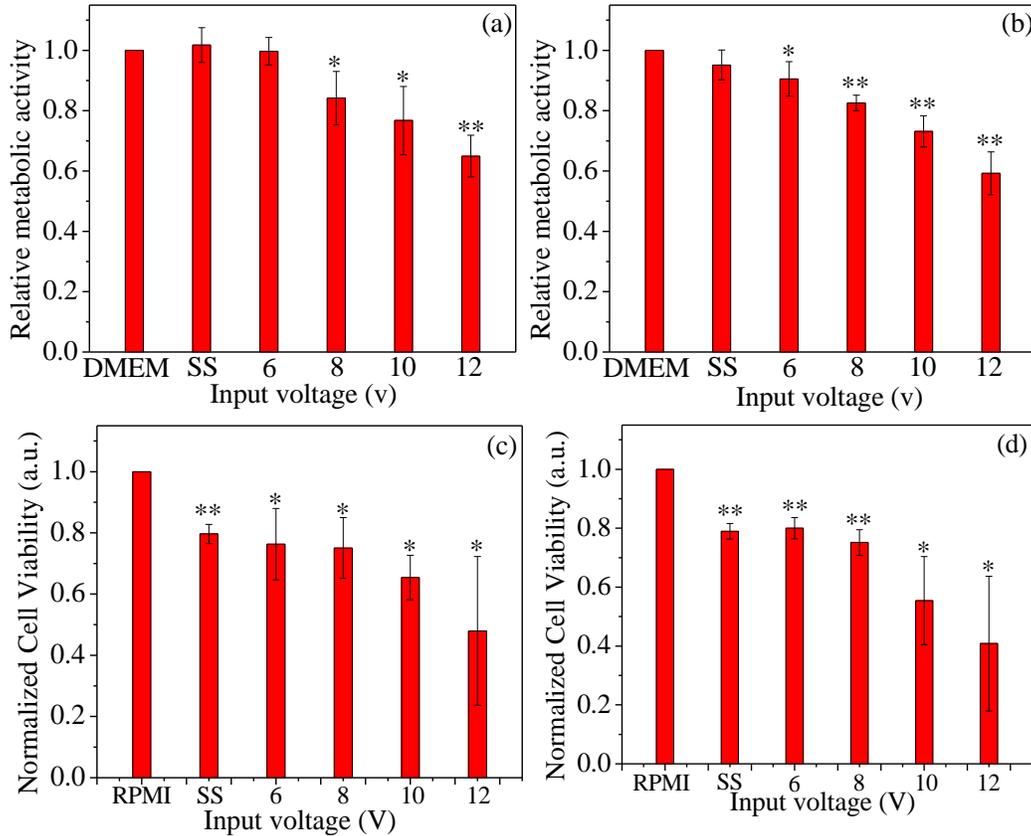

Figure 6. Cell viability of MDA-MB-231 and BxPC-3 treated by He SOP plasma-activated saline solution using copper as lower electrode with input voltage 6, 8, 10, and 12V. 24 (a) and 48 (b) hours' cell viability of MDA-MB-231 cancer cells. 24 (c) and 48 (d) hours' cell viability of BxPC-3 cancer cells. The ratios of surviving cells for each cell line were calculated relative to controls (DMEM/RPMI). Student t-test was performed, and the statistical significance compared to cells present in DMEM/RPMI is indicated as *$p < 0.05$, **$p < 0.01$, ***$p < 0.001$. (n=3)

To investigate the effect of He SOP plasma-activated saline solution and DI water using copper and stainless steel as lower electrodes, the human breast (MDA-MB-231) and human pancreas adenocarcinoma (BxPC-3) cancer cells were treated. Dulbecco's Modified Eagle's Medium (DMEM) and RPMI-1640 Medium were used as the control platform for MDA-MB-231 and BxPC-3, respectively. Cancer cells were cultured in completed DMEM and RPMI media, therefore, we added DMEM and RPMI to cancer cells having no effect (control group), respectively. However, cancer cells will be dead if they are cultured in saline solution and DI water. On the other hand, we compared saline solution and DI water with He SOP plasma-activated saline solution and DI water applied to cancer cells. He SOP plasma-activated saline solution and DI water had more effect on cancer cells than the saline solution and DI water, which means plasma solutions work from He SOP plasma not from saline solution and DI water themselves.

Fig. 6 and Fig. 7 show the effect of He SOP plasma-activated solutions using copper as a lower electrode on both cancer cells. In Fig. 6a-6d, we can see that the cell viability of MDA-MB-231 and BxPC-3 exposed to DMEM/RPMI, saline solution, and saline solution treated by He SOP plasma at input voltage 6, 8, 10, and 12 V for 90 seconds' treatment, incubated for 24 h (Fig. 6a and 6c) and 48 h (Fig. 6b and 6d). We can see that He SOP plasma-activated saline solutions have a stronger effect on the BxPC-3 than that on MDA-MB-231. For both cancer cells, the cell viability decreased with increasing input voltage. Fig. 7a-7d show the cell viability of MDA-MB-231 and BxPC-3 exposed to DMEM/RPMI, DI water, and DI water treated by He SOP plasma at input voltage 6, 8, 10, and 12 V for 90 seconds' treatment, incubated for 24 h (Fig. 7a and 7c) and 48 h (Fig. 7b and 7d). He SOP plasma-activated DI water also has a stronger effect on the BxPC-3 than that on MDA-MB-231. Similar dose-dependent (i.e. increase input voltage) decreases in cell viability were found in Fig. 7. Comparing Fig. 6 with Fig. 7, He SOP plasma-activated DI water has more effect of both cancer cells than that of saline solution.

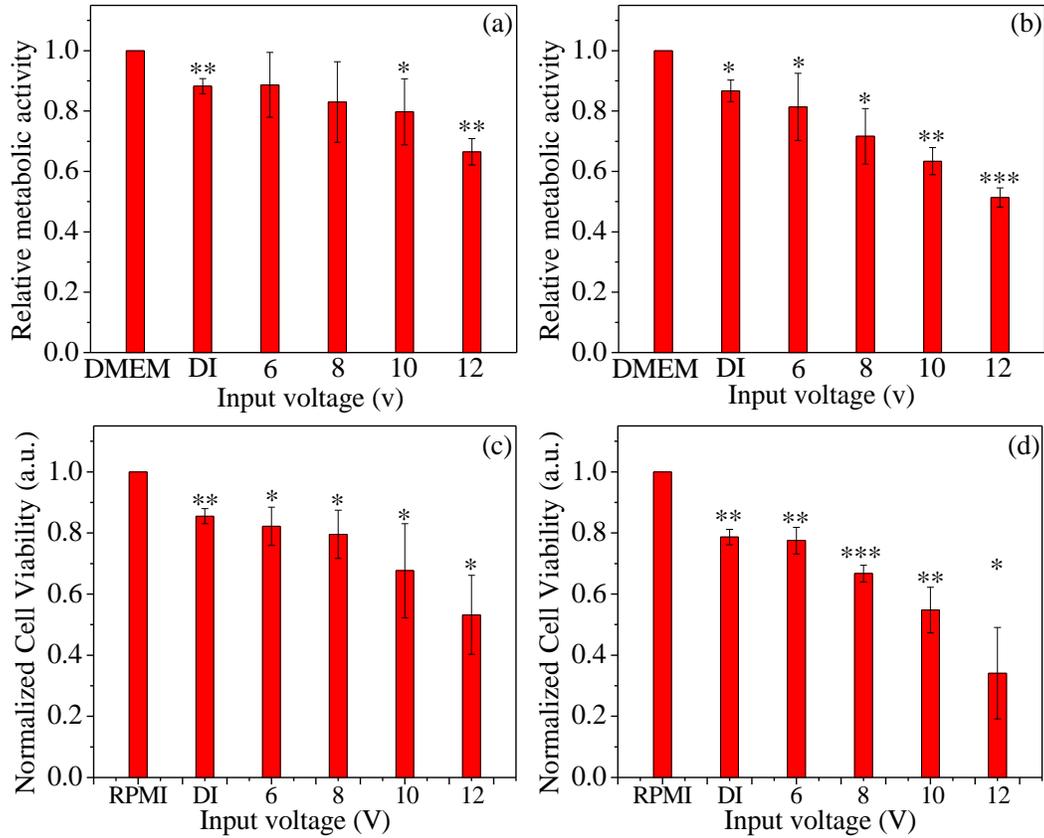

Figure 7. Cell viability of MDA-MB-231 and BxPC-3 treated by He SOP plasma-activated DI water using copper as lower electrode with input voltage 6, 8, 10, and 12V. 24 (a) and 48 (b) hours' cell viability of MDA-MB-231 cancer cells. 24 (c) and 48 (d) hours' cell viability of BxPC-3 cancer cells. The ratios of surviving cells for each cell line were calculated relative to controls (DMEM/RPMI). Student t-test was performed, and the statistical significance compared to cells present in DMEM/RPMI is indicated as $*p < 0.05$, $**p < 0.01$, $***p < 0.001$. (n=3)

*3.5 Cell viability of He SOP plasma-activated saline solution and DI water using stainless steel as lower electrode*

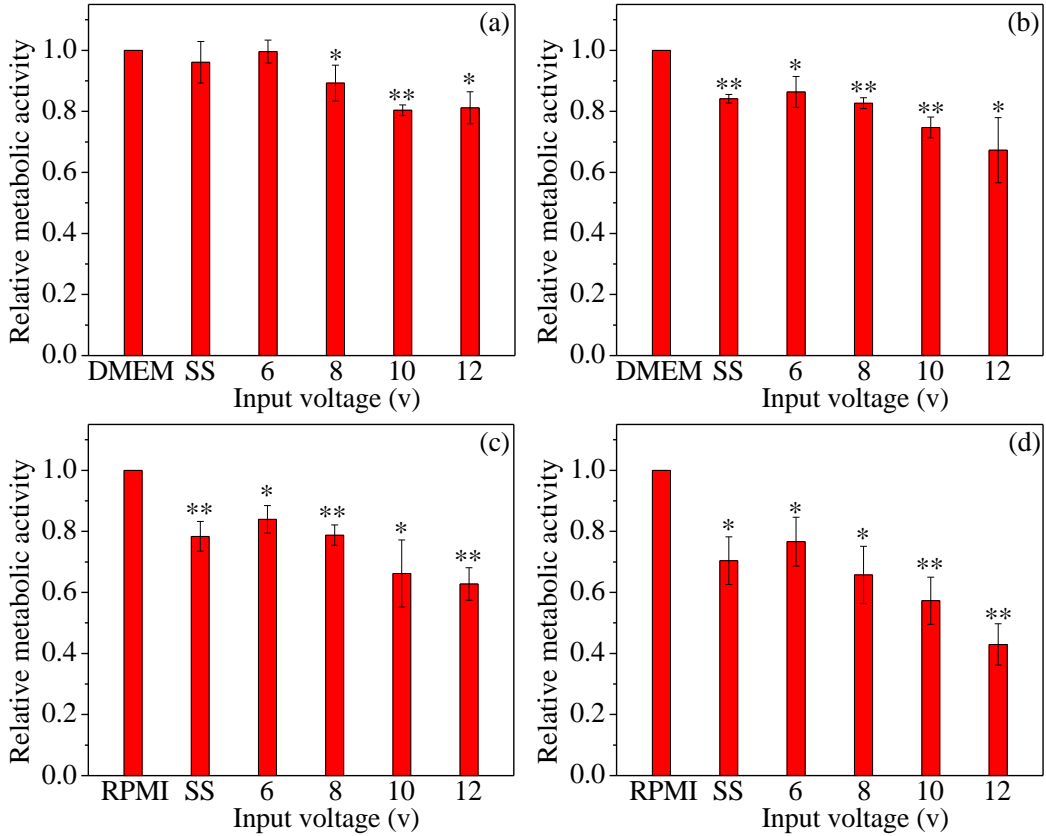

Figure 8. Cell viability of MDA-MB-231 and BxPC-3 treated by He SOP plasma-activated saline solution using stainless steel as lower electrode with input voltage 6, 8, 10, and 12V. 24 (a) and 48 (b) hours' cell viability of MDA-MB-231 cancer cells. 24 (c) and 48 (d) hours' cell viability of BxPC-3 cancer cells. The ratios of surviving cells for each cell line were calculated relative to controls (DMEM/RPMI). Student t-test was performed, and the statistical significance compared to cells present in DMEM/RPMI is indicated as $*p < 0.05$, $**p < 0.01$, $***p < 0.001$. (n=3)

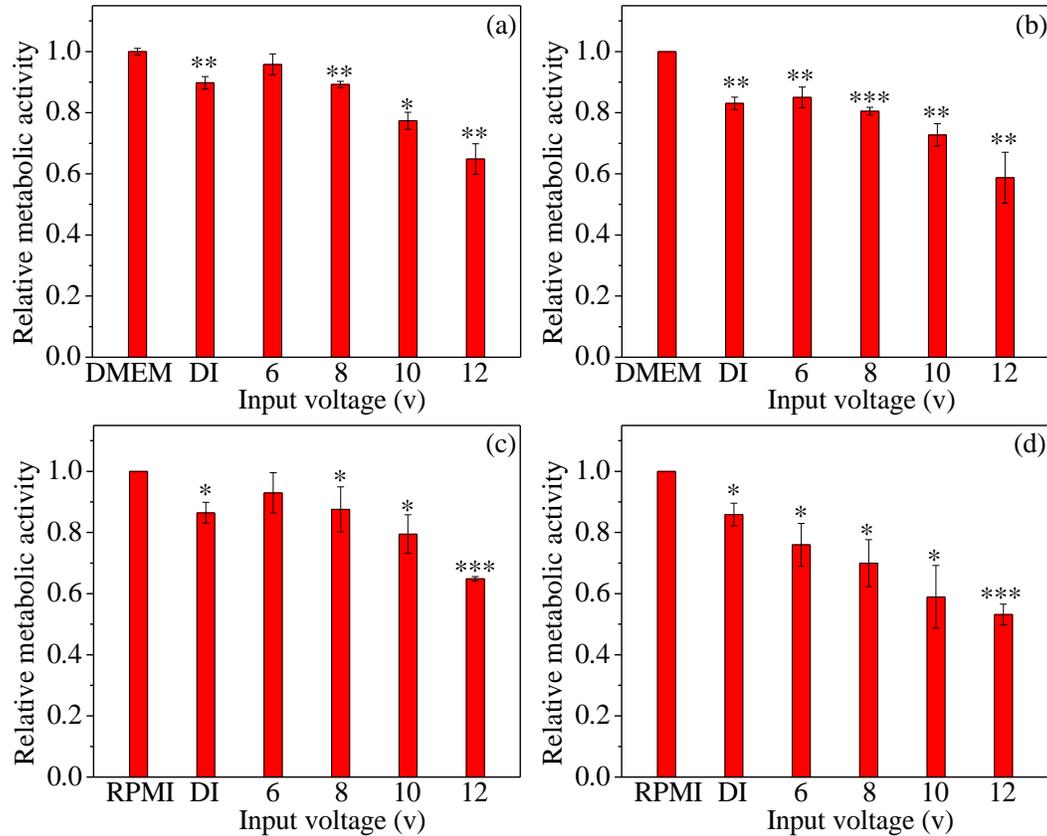

Figure 9. Cell viability of MDA-MB-231 and BxPC-3 treated by He SOP plasma-activated DI water using stainless steel as lower electrode with input voltage 6, 8, 10, and 12V. 24 (a) and 48 (b) hours' cell viability of MDA-MB-231 cancer cells. 24 (c) and 48 (d) hours' cell viability of BxPC-3 cancer cells. The ratios of surviving cells for each cell line were calculated relative to controls (DMEM/RPMI). Student t-test was performed, and the statistical significance compared to cells present in DMEM/RPMI is indicated as *$p < 0.05$, **$p < 0.01$, ***$p < 0.001$. (n=3)

Fig. 8 and Fig. 9 show the effect of He SOP plasma-activated saline solution and DI water using stainless steel as a lower electrode on MDA-MB-231 and BxPC-3 cancer cells. In Fig. 8a-8d, the cell viability of MDA-MB-231 and BxPC-3 exposed to DMEM/RPMI, saline solution and He SOP plasma-activated saline solutions at input voltage 6, 8, 10, and 12 V for 90 seconds' treatment was shown, incubated for 24 h (Fig. 8a and 8c). For MDA-MB-231cancer cells, when incubated for 24 h, cell viability decreased first, then increased at input voltage 10 V. For BxPC-3 cancer cells, cell viability decreases with increasing input voltage, showing a dose-dependent effect. Fig. 9a-9d shows the 24 and 48 hours' cell viability of MDA-MB-231 and BxPC-3 exposed to DMEM/RPMI, DI water, and He SOP plasma-activated DI water at input voltage 6, 8, 10, and 12 V for 90 seconds duration. For both MDA-MB-231 and BxPC-3 cancer cells, cell viability decreases with increasing input voltage. Comparing Fig. 8 and Fig. 9, He SOP plasma-activated DI water has more effect on MDA-MB-231 cancer cells, while He SOP plasma-activated saline solution has more effect on BxPC-3 cancer cells. When comparing copper with stainless steel using as a lower electrode, we can see that He SOP plasma-activated solutions using copper as a lower electrode have more effect on both cancer cells than that of stainless steel.

## 4. Discussions

Atmospheric pressure plasma discharge with self-organized pattern (SOP) is a growing research area, which attracts strong attention of experts in various fields to deeply explore the physical mechanisms behind the self-organization[2,31]. Apart from the fascinating phenomena, the effect of SOP on tumors is among the most important feature of this system. In this paper, we have studied the effect of He SOP plasma on the composition of liquid media at the three discharge modes (I, II, and III), as shown in Fig. 2. Apparently, discharge of modes I, II, and III leads to a large number of hardly predictable chemical reactions, and species generated in reactions can be applied to cancer therapy. Here, we consider He SOP plasma-activated saline solution and DI water because they have the potential to be utilized as oral medicines or to even be paired with other drugs or used as a standalone drug. On the other hand, blood is a body fluid that delivers necessary substances, which is composed of blood cells suspended in blood plasma. Blood plasma constitutes 55% of blood fluid that is mostly water (92% by volume). It is very difficult to directly treat blood by plasma due to its coagulation. Therefore, we can use the strategy of stimulating saline solution and DI water and injecting them into blood around the tumor.

Plasma contains the free radicals, reactive species, energies ions, ultraviolet (UV) radiation, and the transient electric fields inherent with plasma delivery[32-34], which induce apoptosis in cancer cells without adverse effect on the normal cells when administered at the comparable dosage provided[35]. It's well known that ROS can induce apoptosis and necrosis, whereas RNS induces damage to DNA resulting in cell death[36,37]. UV radiation treats water to form ROS at a UV wavelength of 200-280 nm[38]. Therefore, it can be argued that UV photons are not the major He SOP plasma species inducing the production of ROS in solutions in our experimental setup (Fig. 3 and Fig. 4). Radicals and electrons generated during He SOP plasma formation can be either short-lived or long-lived. These radicals or electrons reach a solution and form many complex reactions, which result in the formation of other short- and long-lived radicals or species. Short-lived radicals or species include superoxide ($O_2^-$), nitric oxide (NO), atomic oxygen (O), ozone ($O_3$), hydroxyl radical (OH), singlet delta oxygen (SOD, $O_2(^1\Delta g)$), etc[25,39,40]. Short-lived radicals or species that react to form long-lived species include hydrogen peroxide ($H_2O_2$) and nitrite ($NO_2^-$)[39]. $H_2O_2$ and $NO_2^-$ concentration of He SOP plasma-activated saline solution and DI water were shown in Fig. 5. The concentration of $H_2O_2$ and $NO_2^-$ in saline solution and DI water rises with input voltage. Analyzing cell viability, one can see that He SOP plasma-activated saline solution and DI water increases cancer cell killing efficiency (Fig. 6 – Fig. 9), in line with the linear increase in $H_2O_2$ and $NO_2^-$ concentration. On the other hand, there might be additional antitumor pathways related to $H_2O_2$ and $NO_2^-$. $H_2O_2$ and $NO_2^-$ are believed to produce peroxynitrite ($H_2O_2 + 2NO_2^- - 2HONOO^-$) that is known to be toxic to cancer cells[37]. In our experiment setup, we employed copper and stainless steel as a lower electrode. Thus, there should be a little amount of copper and iron ions in saline solution and DI water after He SOP plasma treatment. Metals ions act as catalysts in the oxidative deterioration of biological macromolecules, and therefore, the toxicities associated with these metals may be due at least in part to oxidative damage. Iron and copper ions exhibit the ability to produce ROS, resulting in lipid peroxidation, DNA damage, and etc[41]. Iron and copper enhance DNA breakage, although copper is more active than iron. This may be the reason for He SOP plasma using copper as lower electrode has more cancer cells killing efficiency than stainless steel.

## 5. Conclusions

In this paper, we present new studies revealing the role and significant potential of He SOP plasma activating saline solution and DI water, capable of efficiently inhibiting the growth and proliferation of the breast cancer MDA-MB-231 and pancreatic cancer cells BxPC-3. Based on the voltage/flow rate characteristics and optical self-organized patterns between electrode and media surface, we have decided on the three quite different discharge modes and demonstrated that the activation under self-organized conditions plays a pivotal role in the synthesis of novel cancer-inhibiting media. Moreover, we have demonstrated that the effect of solution conductivity on discharge modes that are capable of efficiently controlling the ROS and RNS concentrations in the therapeutical media. He SOP plasma using copper as a lower electrode has more cancer cells killing efficiency than stainless steel. Understanding the cancer impact of He SOP plasma discharge modes activating media will help the plasma efficacy in order to tailor them to the therapeutic applications' needs.